\documentclass[12pt,twocolumn]{aastex631}
\usepackage{verbatim, graphicx, physics, hyperref,}

\graphicspath{{./}{figures/}}

\begin{document}

\received{March 19, 2025}
\revised{July 10, 2025}
\accepted{July 11, 2025}
\published{September 4, 2025}
\submitjournal{\apjl}

\title{High Resolution ALMA Data of the Fomalhaut Debris Disk Confirms Apsidal Width Variation}

\author[0000-0002-4985-028X]{Jay S. Chittidi}
\affiliation{Department of Physics \& Astronomy, John Hopkins University, 3400 N. Charles Street, Baltimore, MD 21218}
\affiliation{Department of Astrophysical and Planetary Sciences, University of Colorado, 2000 Colorado Avenue, Boulder, CO 80309, USA}

\author[0000-0001-7891-8143]{Meredith A. MacGregor}
\affiliation{Department of Physics \& Astronomy, John Hopkins University, 3400 N. Charles Street, Baltimore, MD 21218}

\author[0000-0002-4248-5443]{Joshua Bennett Lovell}
\affiliation{Center for Astrophysics, Harvard \& Smithsonian, 60 Garden Street, Cambridge, MA 02138-1516, USA}

\author[0000-0002-5092-6464]{Gaspard Duchene}
\affiliation{Astronomy Department, University of California, Berkeley CA 94720-3411, USA}
\affiliation{Universit\'{e} Grenoble Alpes/CNRS, Institut de Plan\'{e}tologie et d'Astrophysique de Grenoble, 38000 Grenoble, France}

\author[0000-0001-9064-5598]{Mark Wyatt}
\affiliation{Institute of Astronomy, University of Cambridge, Madingley Road, Cambridge CB3 0HA, UK}

\author[0000-0002-6648-2968]{Olja Panic}
\affiliation{School of Physics and Astronomy, University of Leeds, Leeds LS2 9JT, UK}

\author[0000-0002-6221-5360]{Paul Kalas} 
\affiliation{Astronomy Department, University of California, Berkeley CA 94720-3411, USA}
\affiliation{Institute of Astrophysics, FORTH, GR-71110 Heraklion, Greece}
\affiliation{Carl Sagan Center, SETI Institute, 189 Bernardo Avenue, Mountain View, CA 94043, USA }

\author{Margaret Pan}
\affiliation{Center for Astrophysics, Harvard \& Smithsonian, 60 Garden Street, Cambridge, MA 02138-1516, USA}

\author[0000-0002-4803-6200]{A. Meredith Hughes}
\affiliation{Department of Astronomy, Van Vleck Observatory, Wesleyan University, Middletown, CT 06459, USA}

\author[0000-0003-1526-7587]{David J. Wilner} 
\affiliation{Center for Astrophysics, Harvard \& Smithsonian, 60 Garden Street, Cambridge, MA 02138-1516, USA}

\author[0000-0001-6831-7547]{Grant M. Kennedy}
\affiliation{Department of Physics and Centre for Exoplanets and Habitability, University of Warwick, Gibbet Hill Road, Coventry CV4 7AL, UK}

\author[0000-0003-4705-3188]{Luca Matr\`{a}}
\affiliation{School of Physics, Trinity College Dublin, the University of Dublin, College Green, Dublin 2, Ireland}

\author[0000-0002-0176-8973]{Michael P. Fitzgerald}
\affiliation{Department of Physics and Astronomy, UCLA, Los Angeles, CA 90095, USA}

\author[0000-0002-3532-5580]{Kate Y. L. Su}
\affiliation{Space Science Institute, 4750 Walnut Street, Suite 205, Boulder, CO 80301, USA}

\begin{abstract}
We present long-baseline observations of the Fomalhaut outer debris disk at 223~GHz (1.3~mm) from ALMA Cycle 5, which we use along with archival short-baseline observations to produce a 0\farcs57 resolution mosaic of the disk at a sensitivity of 7~$\mu$Jy/bm. We use radial profiles to measure the disk at the ansae and find that the southeast (SE) side of the disk is 4~AU wider than the northwest (NW) side as observed by ALMA. We also find that the peak brightness of the NW ansa is $21\pm1\%$ brighter than the SE ansa. We perform MCMC fits of the ALMA visibilities using two analytical, eccentric disk models. Our results suggest that the model including a dispersion parameter for the proper eccentricity ($\sigma_{e_p}$), which accounts for additional scatter in the eccentricity of individual orbits, is preferred over the model without one.  Such a model implies that self-gravitation, particle collisions, and close-packing could play a role in shaping the overall structure of the Fomalhaut disk as is seen in eccentric planetary rings. Crucially, neither model can reproduce the brightness or
width asymmetry near the NW ansa. No emission from the Intermediate Belt is detected, allowing us to place a 3-$\sigma$ upper limit of 396 $\mu$Jy at 1.3 mm. We also discover a spectral line in archival Cycle 3 data centered at $\nu_{\rm obs}\approx230.25$ GHz at the location of the ``Great Dust Cloud," whose redshift from the expected CO line for Fomalhaut confirms the source is a background galaxy.
\end{abstract}


\section{Introduction}
\label{sec:intro}

Fomalhaut is a nearby \citep[$\sim$7.7~pc,][]{vanLeeuwen2007} A3V star with an estimated age of 440~Myr \citep{Mamajek2012}. Observations from the \textit{Infrared Astronomical Satellite (IRAS)} decades ago revealed a strong IR excess indicative of circumstellar dust, making the system an exciting site to study planetary dynamics and disk evolution \citep{Backman1993}. Low-resolution imaging observations with the Submillimetre Common-User Bolometer Array (SCUBA) at the James Clerk Maxwell Telescope (JCMT) revealed asymmetric emission from the debris disk at the southern ansa \citep{Holland1998}. \citet{Wyatt1999} noted that this could be evidence of an eccentric disk sculpted by an eccentric, hidden planet, resulting in increased thermal emission near the disk pericenter (presumably the southern ansa). \textit{Spitzer} observations presented in \citet{Stapelfeldt2004} showed further evidence of this ``pericenter glow". \textit{Hubble Space Telescope} imaging resolved the outer cold belt \citep[$\sim$140~AU,][]{Kalas2005} and later the appearance of what was initially thought to be a planet but is now thought to be an expanding dust cloud just interior to the disk \citep{Kalas2008, Kenyon2014, Lawler2015,Gaspar2020}. The optical imaging with \textit{Hubble} revealed that the disk was eccentric, and both HST and far-infrared imaging with \textit{Herschel} observed pericenter glow \citep{Acke2012}. Notably, longer-wavelength imaging using facilities such as the Atacama Large Millimeter/submillimeter Array (ALMA) has shown a pronounced brightness enhancement at the northern ansa analogously called ``apocenter glow" \citep{MacGregor2017}. \citet{Pan2016} argue that apocenter glow is due to a surface density enhancement from particles traveling slowest (and hence spending more time) at apocenter, while recent theoretical work done by \citet{Lynch2022} argue that such features also depend on the eccentricity profile of the disk. \textit{JWST} MIRI imaging presented in \citet{Gaspar2023} further revealed the presence of an inner disk of warm grains, a previously undiscovered intermediate belt at $\sim$90~AU just interior to the main belt, and an extended halo of dust from the outer belt also present in HST imaging \citep{Kalas2013}. Recently, \citet{Sommer2025} presented an analysis that demonstrated that the emission from the inner and possibly intermediate belts could be the natural consequences of Poynting-Robertson (PR) drag acting on smaller dust grains from the outer belt. Later, \textit{JWST} NIRCAM imaging presented in \citet{Ygouf2024} placed constraints on planet masses $\leq M_J$ beyond about 8 AU. They found one possible new candidate source (``S7") that future observations will need to search for to verify if it is associated with Fomalhaut. The presence of multiple eccentric debris belts and their multi-wavelength brightness distributions provides an exciting but complex system to study the processes that drive the evolution of planetary systems.

ALMA has revolutionized our understanding of planetesimal disks making many dozens of detections of Kuiper Belt analogs including the Fomalhaut outer belt itself \citep{Sepulveda2019,Marino2019,Marino2022}. \citet{Boley2012} observed the disk in ALMA Cycle 0 and found that the disk edges were sharper than expectations for the system, and appeared consistent with sculpting from an inner and possibly outer planetary mass companion. \citet{MacGregor2017} used ALMA to produce a mosaic map of the disk in Band 6 at 223~GHz, providing high resolution (natural weight beam size of $1\farcs56 \times 1 \farcs15$) and sensitivity (rms of 14~$\mu$Jy/beam) in the sub-mm regime for the disk, and showed that a model which treats the complex eccentricity as the vector sum with a proper and forced component with independent phase parameters was a good match to the observations. \citet{Kennedy2020} found that those same observations showed evidence that the northwest (NW) ansa of the disk was narrower than the southeast (SE) ansa by about 4~AU (measured from the FWHM), and demonstrated that a modified version of the complex eccentricity model that included a dispersion in the forced eccentricity could allow for a narrower NW ansa (proximate to the disk apocenter). 

Here, we present new ALMA observations that reveal the Fomalhaut debris disk at unprecedented resolution at millimeter wavelengths allowing us to more accurately constrain the geometry of the outer belt.  This paper is organized as follows -- in \S\ref{sec:data} we discuss the ALMA datasets, their processing, and our method to correct for the proper motion between the three epochs. In \S\ref{sec:imaging} we present continuum images of the aligned datasets and discuss the \texttt{CLEAN} parameters that were used to produce them. We then present MCMC fit results for two eccentric disk models in \S\ref{sec:model}.  We measure and discuss the widths of the disk at the ansae using radial profiles of the disk images that have been re-gridded to $R-\theta$ space in \S\ref{sec:width}. Specifically, we find that while we can still model the bulk parameters of Fomalhaut's eccentric ring, our models do not provide a good physical interpretation for the disk's asymmetries identified in the higher-resolution data (i.e., the width and brightness differences at the ansae). We also present the detection of a spectral line coincident with the ``Great Dust Cloud" (GDC) source discussed in \citet{Gaspar2023} and \citet{Kennedy2023}, which corroborates its nature as a background galaxy rather than an object associated with the Fomalhaut system. Finally, we discuss our interpretation in the context of alternative descriptions of this system such as those with differences in their radial profile parameterizations and orbital eccentricity distributions of orbital eccentricities, including that of \citet{Lovell2025} analyzing these same observations.

\section{Data}
\label{sec:data}

\begin{table}[]
    \centering
    \begin{tabular}{ccccc}
    \hline
    \hline
    Date & Antennas  & Baselines & PWV & Obs. Time \\
     & & [m] & [mm] & [min]\\
    \hline
    2018 Sept 08 & 45 & 15.1 - 783.5 & 1.3 & 73.1 \\
    & 45 &15.1 - 783.5 &1.2& 72.7 \\
    2018 Sept 17& 45 & 15.1 - 1245.6 &0.4& 78.4 \\
    2018 Sept 23 & 48 & 15.1 - 1397.8 &0.5& 73.1\\
    2018 Sept 26 & 47 &15.1 - 1397.8 &0.8&73.7\\
    \end{tabular}
    \caption{ALMA Observations from Project 2017.1.01043.S}
    \label{tab:obs}
\end{table}

Three epochs of ALMA data in Band 6 were obtained from the archive: a high-resolution pointing on the central star observed in Cycle 2 \citep[][ID\#2013.1.00486.S]{White2017}, a seven-pointing mosaic from Cycle 3 \citep[][ID\#2015.1.00966.S]{MacGregor2017}, and two high-resolution pointings at the disk apses from Cycle 5 (ID\#2017.1.01043.S). The Cycle 5 observations are summarized in Table \ref{tab:obs}. The observations from Cycle 2 were calibrated using ALMA pipeline version 4.3.1, the Cycle 3 data were processed with 4.5.3, and the Cycle 5 data with version 5.1.1-5, all of which are the pipeline versions recommended by ALMA support for each respective data set. The flux calibration and data weights were inspected across the three datasets and were found to be consistent across the three epochs, given their respective baseline coverage, and so no further changes were made to either product. We used two versions of \texttt{CASA} for this work: version 6.5.1-23 was used for all data manipulation tasks such as averaging and visibility model subtraction, while version 6.6.0-20 was used exclusively for imaging due to relevant updates to \texttt{tclean}.

The data were then concatenated into a single {\tt CASA} measurement set. To reduce the data volume, the measurement set was averaged down to 8 channels for the three continuum spectral windows, and 128 channels for the spectral window centered on the 230.538~GHz CO line. We then time-averaged the data to 30-second intervals.

Self-calibration was attempted in order to correct for phase offsets introduced from the star's proper motion \cite[328.95, -164.67~ mas~yr$^{-1}$, comparable to the net beam size across the three epochs,][]{vanLeeuwen2007}, but was unsuccessful due to low signal-to-noise. Instead, we used the \texttt{CASA} implementation of \texttt{uvmodelfit} \citep{Marti-Vidal2014} to fit for the position of the star for each unique observation and pointing (a total of 52 fits). We then used the last observation from the Cycle 3 data as the reference position of the star, $\alpha= \rm 22^{h}57^{m}39.450801$, $\delta= \rm -29^{\circ}37$'$22 \farcs 69400$, and used the \texttt{fixplanets} and \texttt{fixvis} functions to manually correct the phase center and UVW baseline positions to the reference stellar position. To validate the proper motion correction, we imaged each of the 52 separate pointings/observations, and analyzed the stellar position in the images for each set of pointings. We found that the stellar centroid position was precise to two pixels ($\sim 0\farcs1$, or about 1/5 of the synthesized beam of the image in Section \ref{sec:imaging}). The scatter is dominated by the star coinciding with the first null in the primary beam for the NW ansa pointing from the seven-pointing, Cycle 3 data. The resulting measurement set is proper motion-corrected for any emission comoving with the star and debris disk, but with smeared emission for background sources.

\section{Imaging}
\label{sec:imaging}

\begin{figure*}
\centering
\includegraphics[width=\textwidth]{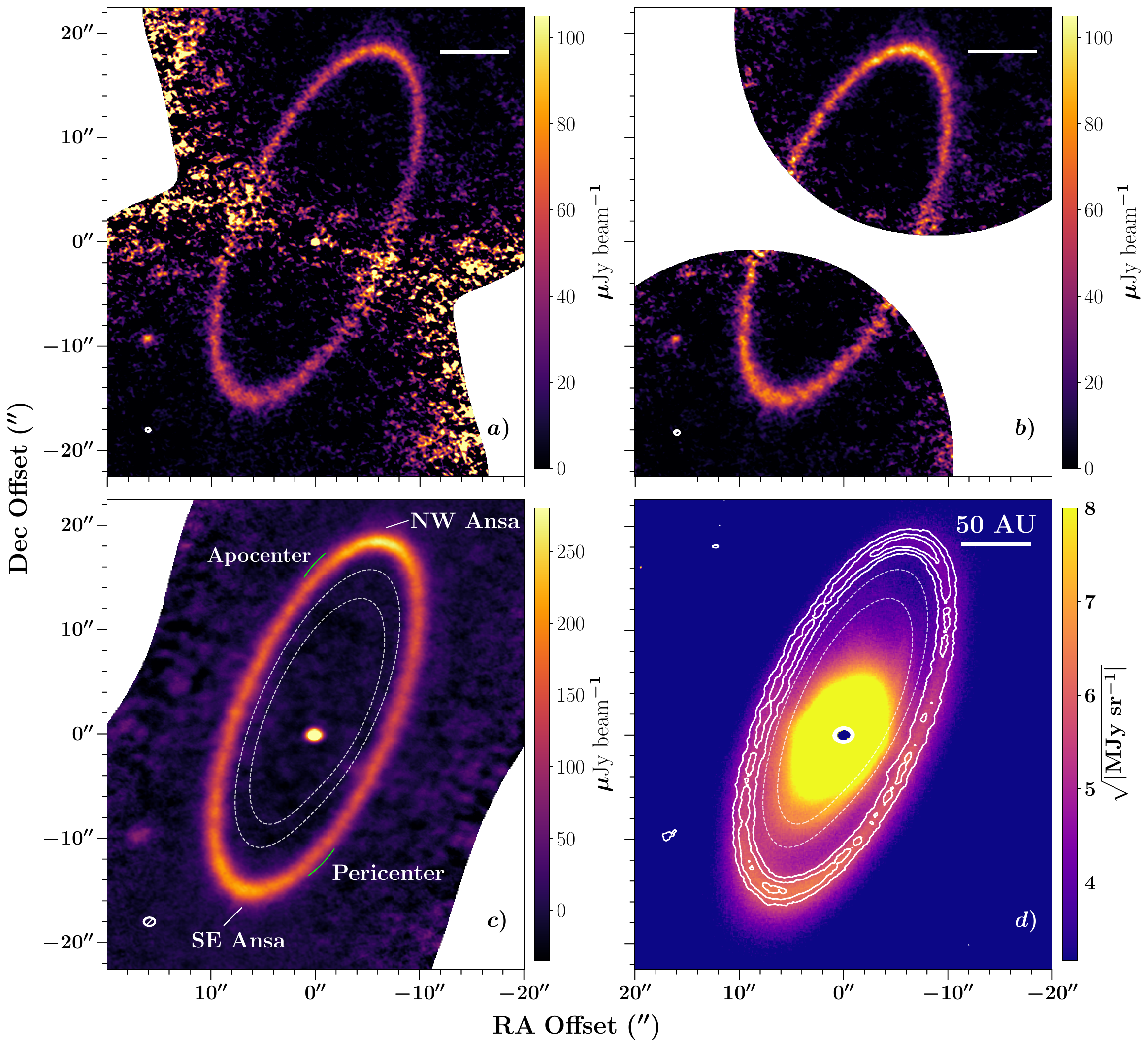}
\caption{\textit{Top Left:} A mosaic of the Cycle 5 (C5) long baseline data. The data were imaged to the 5$\%$ gain level to highlight the detection of the star, which sits at the edge of the primary beam between the NW and SE ansae pointings. The naturally-weighted beam size is $0\farcs50 \times 0\farcs39$ (indicated by the ellipse in the lower left corner of each panel) and the rms noise near the pointing centers are 7 $\mu$Jy/bm. The axes are the stellocentric offset in arcseconds and the image is North-aligned. \textit{Top Right:} A mosaic of the NW and SE ansae pointings using the coaligned shorter baseline Cycle 3 (C3) data as well as the longer baseline C5 data. The naturally-weighted beam size is slightly larger, 0$\farcs57 \times 0\farcs44$, and the rms is also 7 $\mu$Jy/bm. \textit{Bottom Left:} A mosaic of all seven pointings of ALMA Band 6 observations from C2, C3, and C5. The naturally-weighted beam size is $1\farcs09 \times 0\farcs81$ and the rms noise near the pointing centers at the ansae are 7 $\mu$Jy/bm, but decreases to 12 $\mu$Jy/bm towards the Intermediate Belt (denoted by the dashed white curves, see Section \ref{sec:ib}). The lack of long baseline observations at the intermediate pointings leads to a beam size that is larger than the naturally-weighted beam corresponding to just the NW and SE ansae data as in the top right. \textit{Bottom Right:} JWST 25.5 $\mu$m image from \citet{Gaspar2023} with overlaid contours from the bottom right ALMA mosaic corresponding to the 10, 20, and 30-$\sigma$ flux levels. The image has been scaled to highlight the intermediate and inner disks and the position of the contours have been corrected for the proper motion between the two imaged epochs. The fainter dashed curves denote the Intermediate Belt boundaries as in the bottom left.}
\label{fig:mosaic}
\end{figure*}

The data were imaged together using the multi-frequency synthesis task in \texttt{tclean} implemented in \texttt{CASA} v.6.6.0-20 in three ways: first we highlight just the NW and SE ansae pointings from \textit{only} the Cycle 5 data, then we evaluate those same two pointings but with the shorter-baseline Cycle 3 data to highlight the improvement to the total flux, and lastly we mosaic \textit{all} of the available data in Band 6 to produce a single image. We opted to use the \texttt{multiscale} deconvolution method with the parameter \texttt{scales=[0,10,20]}, which allows for Gaussian sources in the \texttt{CLEAN} model map equal to the number of pixels specified by \texttt{scales} (0 is equivalent to typical \texttt{CLEAN} point source). We found that this resulted in a \texttt{CLEAN} model that looked more like a continuous, resolved disk rather than a collection of point sources as produced with the \texttt{hogbom} algorithm. Since the mosaics were constructed with uneven sensitivity across the pointings, and the disk extends across any individual pointing, all of the images presented here are primary-beam corrected. The image pixel scale was set so that the synthesized beam major axis covered approximately 10 pixels given the selected data for each image, and then the choice of \texttt{scales} was determined experimentally so that the largest scale was smaller than the apparent width of the disk. In addition, we used an elliptical \texttt{CLEAN} mask that encompassed the outer disk and star with parameters: $\alpha= \rm 22^{h}57^{m}39.443347$, $\delta= \rm -29^{\circ}37$'$21 \farcs 12931$, $a=25''$, $b=12.5''$, and $\theta=338^{\circ}$. For the image of just the Cycle 5 data, we imaged the data to the 5\% gain level to include the strongly-detected star, but used a primary beam mask at the 30\% gain level in lieu of the elliptical mask discussed above to avoid creating anomalously bright \texttt{CLEAN} sources at the edge of the primary beam. The combined Cycle 3 and 5 image, as well as the full mosaic were imaged to the 20$\%$ gain level. All images were \texttt{CLEAN}ed to 3$\times$ the rms noise level. We present naturally-weighted images in this paper in Fig.~\ref{fig:mosaic}.
Briggs-weighted images are presented and analyzed in \citet{Lovell2025}.

In Figure \ref{fig:mosaic}a, we show the image of the Cycle 5 data, while in \ref{fig:mosaic}b we show the combined Cycle 3 and Cycle 5 data. The rms noise near the phase centers is about 7 $\mu$Jy, and the naturally-weighted beam sizes are $0\farcs50 \times 0\farcs39$ and $0\farcs57 \times 0\farcs44$, respectively. In Figure \ref{fig:mosaic}c, we present the mosaic image of \textit{all} of the available data. The synthesized beam size is $1\farcs09 \times 0\farcs81$ and the rms noise near the pointing centers is also 7 $\mu$Jy. The  different resolutions between the Cycle 2 and Cycle 5 data and the lack of high-resolution observations for the four ``intermediate" pointings adjacent to the disk minor axis result in image artifacts that increase the rms noise near the star. In addition, the inconsistent resolution effectively results in a restoring beam size that is larger than the naturally-weighted beam size for the data at the disk ansae (Figure \ref{fig:mosaic}b), and so in the analysis that follows, we consider both the full and partial mosaics depending on our use case. The disk modeling in Section \ref{sec:model} bypasses the non-linearity inherent to \texttt{CLEAN} and the above artifacts altogether by directly comparing the data and model visibilities.

\subsection{Image Analysis}
\label{sec:imageanalysis}

\begin{figure*}
    \centering
    \includegraphics[width=\textwidth]{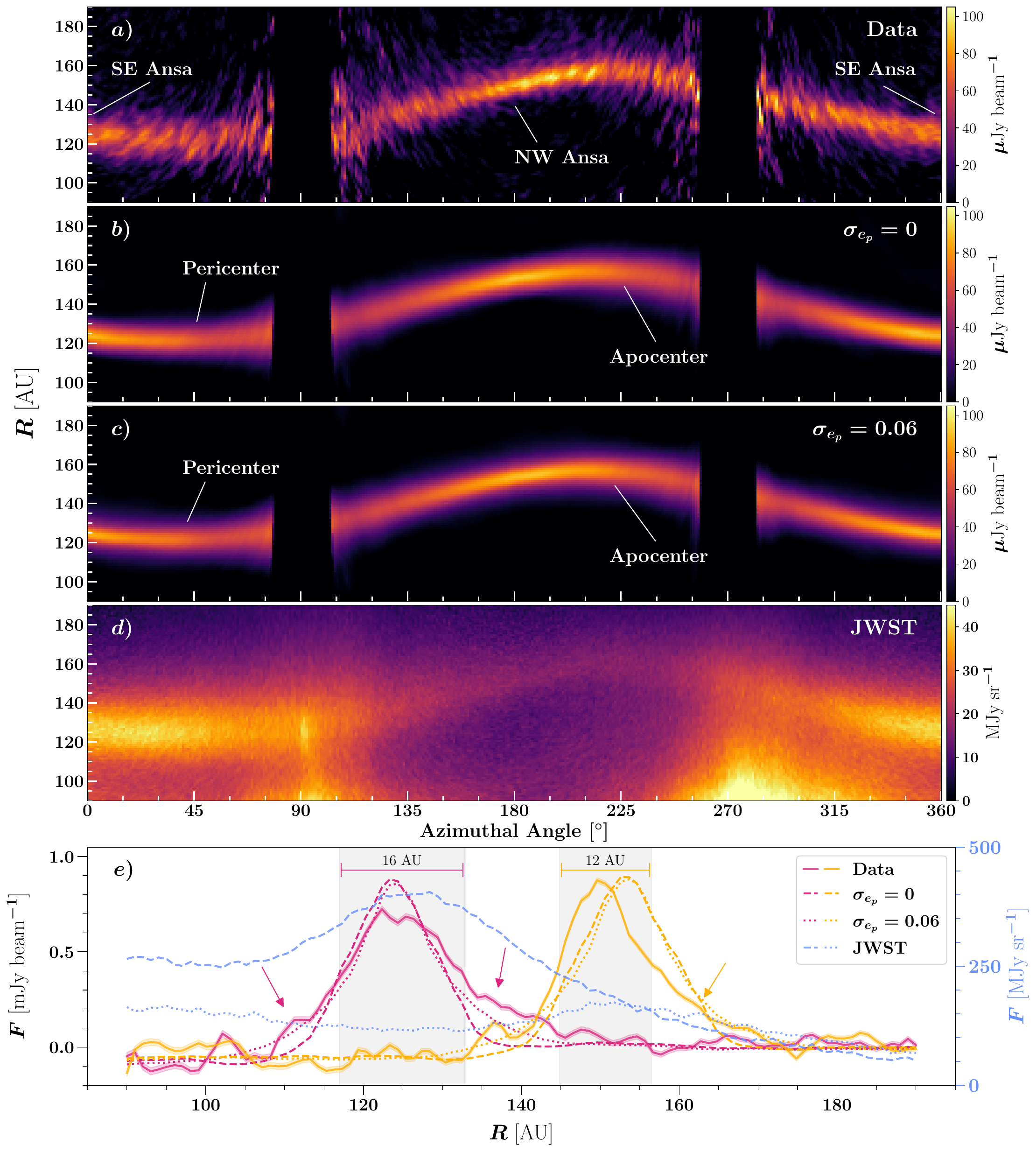}
    \caption{\textit{First Row:} A deprojected map of the circumstellar radius vs. azimuthal angle of the Fomalhaut outer disk produced using the mosaic in Figure \ref{fig:mosaic}b. The angular coordinates are with respect to the disk position angle of 335.84$^{\circ}$. \textit{Second Row:} A radius vs. azimuth map for the best-fit model without a dispersion in eccentricity, after imaging with \texttt{CASA}. \textit{Third Row:} Same as the second, but for the best-fit model with a dispersion in eccentricity.
    \textit{Fourth Row:} A radius vs. azimuth map of the 25.5 $\mu$m JWST image of the disk. Unlike the ALMA images, the NW ansa is significantly dimmer than the SE ansa.
    \textit{Fifth Row:} Radial profiles produced using 10$^{\circ}$ radial cuts centered at 0$^{\circ}$ (SE ansa near apocenter, red) and 180$^{\circ}$ (NW ansa near apocenter, yellow) from the above maps. For the ALMA data, the $21\pm1\%$ brightness asymmetry between the ansae is strongly apparent. The labeled shaded regions are the measured FWHM of the data. ``Shoulders" of emission at 20 AU interior and exterior to the peak flux in the SE ansa profile, and exterior to NW ansa profile are denoted by the colored arrows. The radius of peak emission differ between the data and the models, which is discussed in the text. }
    \label{fig:deproj}
\end{figure*}

In order to analyze azimuthal variations in the width of the disk, we deproject the disk image from Figure \ref{fig:mosaic}b (which has a smaller beam size than the full mosaic) using the best-fit inclination and position angle from the MCMC results presented later in Table \ref{tab:modelfits}, $i=66.5^{\circ}$ and PA$=335.84^{\circ}$. We then resample the image onto a grid of the circumstellar radius and azimuthal angle by averaging the flux of pixels corresponding to the same $dR-d\theta$ bin, presented in Figure \ref{fig:deproj}a. The angular coordinates are measured relative to the disk position angle (i.e. the sky plane), so $0^{\circ}$ and $360^{\circ}$ correspond to $335.84^{\circ}$ on the sky. In this view, the true pericenter and apocenter appear to occur at $\approx$45$^{\circ}$, consistent with the results presented later for the $\omega_f$ parameter from the MCMC fits (though we note that the \textit{apparent} apsides from these plots alone are obscured by both the disk scale height and change in resolution due to the beam position angle). The projection leads to increased artifacts at the disk minor axes, though this region is masked by our choice of the primary beam limit in the two-pointing mosaic.

We then use the deprojected map to create 1D radial profiles using a 10$^{\circ}$ wedge centered at the disk ansae (0$^{\circ}$ and 180$^{\circ}$ along the x-axis) in Figure \ref{fig:deproj}e. The $3-\sigma$ uncertainty in the ALMA data profiles are plotted by using the 7$\mu$Jy beam$^{-1}$ measured in the image and as a function of number of binned pixels relative to the number of pixels in the beam area: $\sigma_{profile}=$7$\mu$Jy$\sqrt{N_{pix/beam}/N_{pix/bin}}$. We compare the peak-to-peak brightness ratio between the radial profiles and measure a $21\pm1\%$ brightness enhancement at the NW ansa, a larger difference than reported in \cite{MacGregor2017}.  This likely results from the low resolution of the 2017 observations, which do not fully resolve the disk width.  Considering just the SE ansa of the disk, the peak brightness is $0.21\pm0.02$ and $0.25\pm0.03$~mJy~arcsec$^{-2}$ in the 2017 data and the new data presented here (Figure 1a), respectively, consistent within the mutual uncertainties.  The difference in peak brightness is more noticeable at the NW ansa where the peaks are $0.23\pm0.02$ and $0.30\pm0.03$~mJy~arcsec$^{-2}$, respectively.  However, the NW ansa is where the resolution difference between the two datasets is most significant given the narrow disk width.  We compute the integrated flux to be $0.23\pm 0.02$~mJy and $0.27\pm 0.03$~mJy, respectively, to account for this resolution difference and note that these values also overlap within the mutual uncertainties.

We compute the width of the disk at each ansa by measuring the full width at half maximum (FWHM) of the radial profiles. The FWHM is 16~AU for the SE ansa profile, and 12~AU for NW ansa profile, resulting in a width difference of 4~AU. The widths are denoted as the highlighted regions in Figure \ref{fig:deproj}e. \citet{Kennedy2020} also measured a 4~AU FWHM width difference, although the respective ansae also each appeared 4 AU wider at lower resolution. 

The radial profiles analyzed here also have a complex shape. We identify the presence of ``shoulders" of emission located at about 15-20~AU interior and exterior to the emission peaks (see the colored arrows in Figure \ref{fig:deproj}e). The inner edge of the apocenter ansa radial profile is also sharper than the outer edge or the pericenter ansa profile. We assess how these extended features influence the profile shape by measuring the full width at 20$\%$ maximum. At this lower flux threshold, the SE ansa width is 29~AU and the NW ansa width is 21~AU, or twice the width difference as before. This suggests that these low surface brightness features are significant.  In the subsequent modeling section, we consider two eccentric disk models. Of key concern is whether either of these models are able to reproduce the 21$\%$ brightness enhancement at apocenter and the approximately 4~AU difference in width identified in the data.

\section{Eccentric Disk Models}
\label{sec:model}

The underlying orbital parameters that describe the surface brightness distribution are modeled using a particle-based approach investigated in \citet{MacGregor2017} and \citet{Kennedy2020} \citep[and more recently for the HD 53143 disk in][]{MacGregor2022}. We consider two disk models with complex forced and proper eccentricities, $e_f$ and $e_p$. The first model most closely mirrors the model initially studied in \citet{MacGregor2017} while the second model includes an additional free parameter for a Gaussian dispersion in the proper eccentricity, which \citet{Kennedy2020} demonstrated could allow for a narrower apocenter than pericenter.

For both models, the mean longitude, $l$, and argument of periastron, $\omega_p$, of orbiting particles are drawn from a uniform distribution between 0 and 2$\pi$. Then, the particles populate the complex eccentricity plane defined by three free parameters: the forced eccentricity, $e_f$, the forced argument of periastron, $\omega_f$, and the proper eccentricity, $e_p$. For the second model, the proper eccentricity is drawn from a normal distribution whose mean is $e_p$ and standard deviation is $\sigma_{e_p}$. To avoid non-physical parameters, we take the absolute value of $e_p$ so $e_p>0$, and redraw particles from the distribution when $e_p>1$. We then solve Kepler's Equation for the true anomaly, $f$, with the \texttt{kepler} code \citep{ForemanMackey2021}.

Next, the semi-major axis of each particle is drawn from a uniform distribution defined between $a$--$\Delta a /2$ and $a$+$\Delta a /2$, where $a$ and $\Delta a$ are free parameters. Then, the radial position of each particle is solved by the equation
\begin{equation}
    r = \frac{a (1 - e^2)}{1 + e\,\mathrm{cos}(f)}
\end{equation}
where $f$ is the true anomaly.

\noindent The particles are given a height, $z$, about the disk midplane and are drawn from an exponential distribution defined by a single free parameter scale height, $h$, such that $z=h/r$. We also account for the disk geometry, fitting for an inclination, $i$, and position angle, PA, defined North to East.  RA and Dec offsets (positive in the North and East directions, respectively) account for any global pointing offset. 

To create an image, we bin the particles onto a 2D spatial grid (histogram) whose values are then scaled by $r^{-0.5}$ in order to simulate a temperature profile.  The total disk flux is normalized such that $F_{\rm belt} = \int I_{\nu}d\Omega$ and a point source representing the star is added with flux $F_{\rm star}$. To simplify comparison, we ensure that both models have 12 free parameters by fixing the stellar flux to $F_{\rm star}$=0.735 mJy for the second model with $\sigma_{e_p}>0$. This lower value for the stellar flux was based on an early fit to the data, but this is independent of the parameters for the disk and should not impact the best-fit parameters. However, when we compute the Bayesian Information Criterion (BIC) using the median parameters from the \texttt{emcee} posteriors, we use the stellar flux from the first model for an even comparison.

\citet{Kennedy2020} and \citet{Lovell2023} demonstrated that line density models like the one considered here need a sufficient number of particles in order to reduce the shot-noise associated with the randomly-generated model and to effectively sample the surface density distribution under the beam. \citet{Kennedy2020} used $10^7$ particles to model the C3 data with a dispersion parameter, which we estimate resulted in a 0.2$\%$ model-induced error based on the shot-noise analysis presented in Figure A1 of \citet{Lovell2023}. We use that same analysis to determine that we would need $\mathcal{O}$($10^8$) particles in order to achieve a similar level of shot-noise at the C5 resolution of 0\farcs5. We use \texttt{galario} \citep{Tazzari2018} to sample the model images into visibilities to compare with the ALMA data. For each unique observation ID, spectral window, and field, the model image is offset to the correct relative pointing before comparing with the visibilities, and thus the net $\chi^2$ for each model image is the sum of the $\chi^2$ values for each unique offset. The parameter space is explored with the Markov Chain Monte Carlo (MCMC) package \texttt{emcee} \citep{ForemanMackey2013} using 80 walkers and 18,027 steps for the first model and 22,199 steps for the second model. We assess the models as having converged when the chains were run for at least 50 times the longest auto-correlation time for any of the parameters.

Analytic models like these have limitations and often fail to completely model complex systems.  For example, the models we employ do not account for either density enhancements or disk broadening as a function of true anomaly.  The Fomalhaut debris disk is likely dynamically influenced by at least one planet \cite[e.g.,][]{Boley2012}, so a complete model of the system would require use of N-body simulations.  However, as noted above, this analytic model is computationally intensive.  Attempting to fold an N-body simulation into this MCMC framework would take a prohibitive amount of computing resources.  As a result, analytic models are extremely useful for understanding disk geometry and are widely used throughout the literature.  Some N-body simulations are included in the companion paper to this article \citep{Lovell2025}, and we defer more complicated modeling to future work.

\subsection{Modeling Results}
\begin{figure*}
\centering
\includegraphics[width=\textwidth]{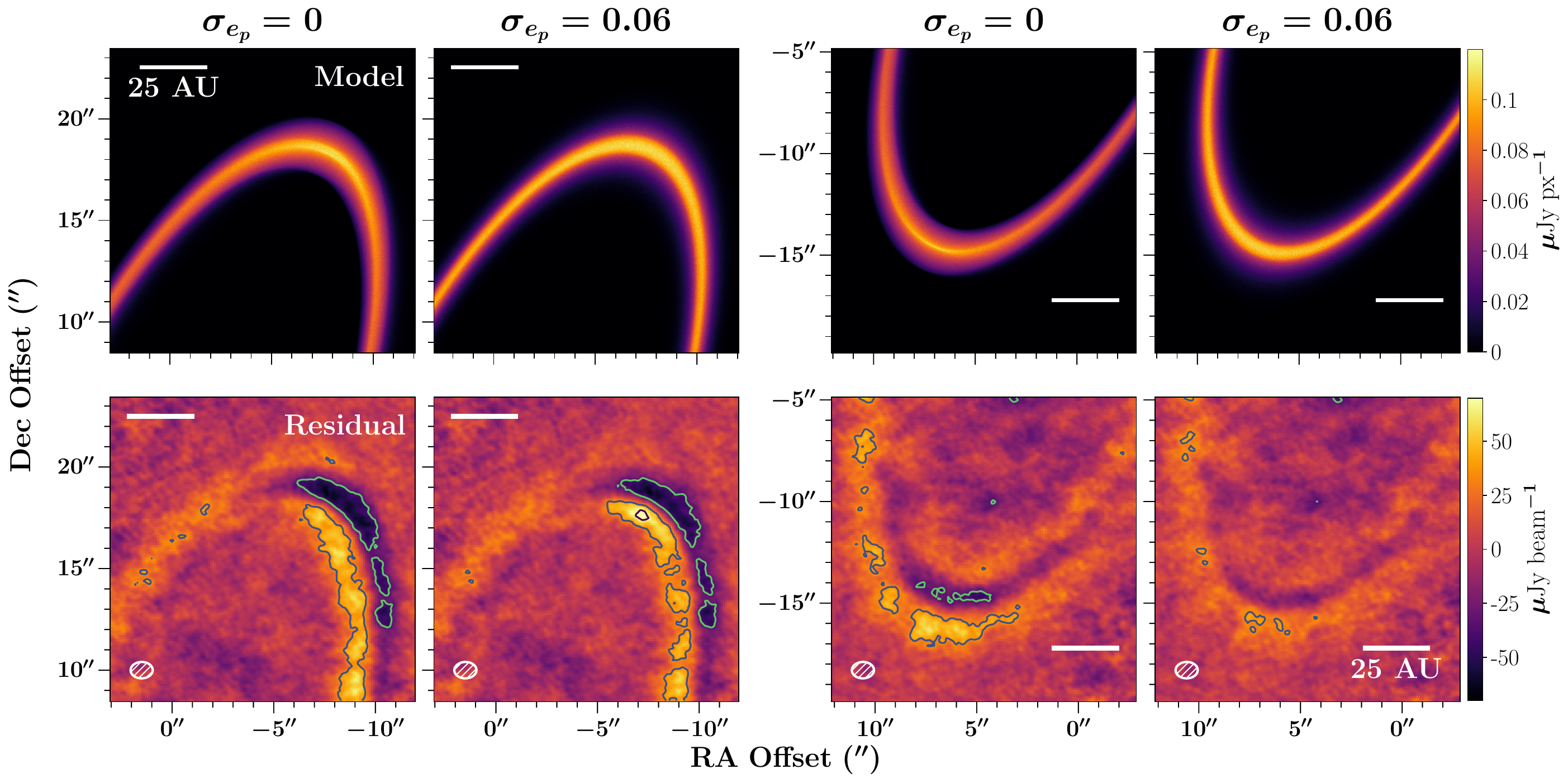}
\caption{\textit{Top Row:} Apocenter and pericenter plots of the best fit eccentric disk models without (first column) and with (second column) a free parameter describing a Gaussian dispersion in the proper eccentricity for the particles (see Table \ref{tab:modelfits}). These images are full-resolution, i.e. not convolved with the synthesized beam, in order to highlight the underlying differences in intensity. \textit{Bottom Row:} Residual between the data and model with contours highlighting -10, -5, 5, and 10-$\sigma$ noise level. These were imaged with \texttt{tclean} with the same parameters as in Figure \ref{fig:mosaic}c. The ellipses indicate the size of the ALMA beam.}
\label{fig:modres}
\end{figure*}

We present the median posterior parameters for both models in Table~\ref{tab:modelfits}, and in Figure~\ref{fig:modres} we show the full-resolution (i.e., not \texttt{CLEAN}'ed or convolved with the synthesized beam) model images at the disk ansae.
With these, we also present the residuals that have been repackaged into a \texttt{CASA} measurement set and imaged with the same \texttt{tclean} parameters as the full-mosaic presented in Figure \ref{fig:mosaic}c.

The nominal disk semi-major axis, inclination, position angle, and scale height for both models are identical and are in good agreement with the analogous ``Uniform simple" and ``Uniform full" model fits presented in Table 1 of \citet{Kennedy2020}. The forced and proper eccentricities for the two models presented here are also in good agreement within their parameter uncertainties, but they do differ slightly from the results in \citet{Kennedy2020}. Here, we find that for the model without dispersion, the forced eccentricity is about 0.02 higher and the proper eccentricity is about 0.03 lower than the \citet{Kennedy2020} results. For the model with dispersion, the results are more in agreement, but the $\sigma_{e_p}$ parameter is about 0.03 lower in this work.
This could be due to a combination of the choice of the uniform distribution of semi-major axes and the ``shoulders". As seen in Figure~\ref{fig:deproj}e, the peak emission for both models near the apocenter side of the disk occurs about 4~AU further out than the data, towards the exterior ``shoulder." The eccentricity parameters and choice of distribution could both affect where this peak occurs, and may explain why our results appear to differ from the literature and the data. Lastly, the model with dispersion has a slightly lower argument of periastron that is comparable to the $41^{\circ} \pm 1 ^{\circ}$ constraint in \citet{Kennedy2020}. The difference in values between our two models does not appear to be statistically significant.

The most notable differences between the two models in this work are the disk width and the total disk flux. The inclusion of the $\sigma_{e_{p}}$ parameter was expected to reduce the disk width such that the expression $\sqrt{(\Delta a)^2 + (a \sigma_{e_{p}})^2}$ should correspond to the disk width \textit{without} the dispersion parameter. Indeed, we find that the expression would yield an effective disk width of 10.70~AU, in excellent agreement with the 10.88~AU result for the model without the dispersion parameter. \citet{Kennedy2020} reported a best-fit dispersion parameter of 0.09, higher than we find here, but this is likely related to the differences noted above. The model with dispersion has a total disk flux about 1~mJy brighter than the other model, though this difference is only at the 2-$\sigma$ significance level. This may be influenced by the ``shoulders" observed in the radial profiles, since the model with dispersion has a wider profile that overlaps with these features while the model without the dispersion does not. \citet{MacGregor2017} reported a total disk flux of 24.7 $\pm$ 0.1~mJy, higher than either model here, either because the ``shoulders" are unresolved in those observations or the inclusion of the long-baseline visibilities weigh down the total flux in the model fits in this work.  Notably, neither model is able to reproduce the 21\% flux difference between the ansae, instead producing roughly equal peaks. 

We compute the BIC for our two models:
\begin{equation}
    \texttt{BIC} = k \mathrm{ln}(n_{\mathrm{vis}}) + \chi^2
\end{equation}
where $k$ is the number of parameters (12 for both models), $n_{\rm vis}$ is the total number of visibilities (real and imaginary, $>8.6\times 10^6$). We include the BIC values in Table \ref{tab:modelfits}. The model with the lowest BIC is the preferred model, with differences of $>10$ indicating that the lower BIC is very-strongly preferred \citep{Kass1995}. We find that the model with $\sigma_{e_p}$ is very-strongly preferred over the model without, in agreement with similar findings in \citet{Kennedy2020}.

\begin{table}
    \centering
        \caption{Eccentric Disk Model Posteriors}
    \begin{tabular}{ccc}
    \hline
    \hline
    Parameter & Model Without $\sigma_{e_p}$ & Model With $\sigma_{e_p}$ \\
    \hline
    $a$ [AU] & $139.48^{+0.39}_{-0.39}$ & $139.48^{+0.43}_{-0.42}$ \\
    $\Delta a$ [AU] & $10.88^{+2.33}_{-2.95}$ & $6.66^{+2.06}_{-2.62}$\\
    $F_{\mathrm{disk}}$ [mJy] & $21.95^{+0.54}_{-0.53}$ & $22.99^{+0.72}_{-0.68}$ \\
    $F_{\mathrm{star}}$ [mJy] & $0.77^{+0.01}_{-0.01}$ & - \\
    $i\; [^{\circ}]$ & $66.50^{+0.11}_{-0.10}$ & $66.50^{+0.11}_{-0.11}$ \\
    PA $[^{\circ}]$ & $335.84^{+0.11}_{-0.12}$ & $335.84^{+0.11}_{-0.11}$ \\
    $e_f$ & $0.15^{+0.01}_{-0.01}$ & $0.14^{+0.01}_{-0.01}$ \\
    $e_p$ & $0.03^{+0.01}_{-0.01}$ & $0.02^{+0.01}_{-0.01}$ \\
    $\sigma_{e_p}$ & - & $0.06^{+0.01}_{-0.01}$ \\
    $\omega_f$ [$^{\circ}$] & $45.94^{+1.38}_{-1.51}$ & $41.89^{+2.94}_{-4.18}$\\
    RA$_{\rm off}$ [''] & $0.08^{+0.01}_{-0.01}$ & $0.08^{+0.01}_{-0.01}$\\
    Dec$_{\rm off}$ [''] & $-0.05^{+0.01}_{-0.01}$ & $-0.05^{+0.01}_{-0.01}$\\
    $h$ & $0.01^{+0.01}_{-0.01}$ & $0.02^{+0.01}_{-0.01}$ \\
    \hline
    BIC & 27815115 & 27811668\\
    \hline
    \end{tabular}
    \label{tab:modelfits}
\end{table}

\section{Discussion}
\label{sec:discussion}

\subsection{Variation of the Disk Width}
\label{sec:width}

\begin{table}
    \centering
        \caption{Disk Ansae FWHM Measurments}
    \begin{tabular}{ccc}
    \hline
    \hline
    Dataset & SE Ansa & NW Ansa \\
    & [AU] & [AU] \\
    \hline
    ALMA Band 6 & 16 & 12 \\
    $\sigma_{e_p}$ = 0 & 12 & 13 \\
    $\sigma_{e_p}$ = 0.06 & 11 & 12 \\
    \hline
    \end{tabular}
    \label{tab:widths}
\end{table}

We generate additional radial maps for the two best-fit models (these are imaged with \texttt{CLEAN} using the same parameters as the two-pointing mosaic) and for the 25.5~$\mu$m JWST MIRI image using the same technique presented in Section \ref{sec:imageanalysis}. We compare radial profile cuts with the data in Figure \ref{fig:deproj}e. Rather critically, neither model investigated in this work seems to be able to reproduce the 21$\%$ brightness asymmetry or the 4~AU width difference. The peak flux in the model profiles are nearly the same at each ansae, and the measured FWHM appears to be 1~AU wider at apocenter ansa, the complete opposite of the trend in the data. These measured widths are presented in Table \ref{tab:widths}. 

The model radial profiles also elucidate the residuals present in Figure \ref{fig:modres} -- these best-fit models have their peak emission occurring at slightly lower radii near pericenter and noticeably higher radii near apocenter compared to the data. This may be due to our choice of a uniform distribution for the disk particle semi-major axes. For example, a different prescription, such as a Gaussian, could alter where the peak brightness occurs. Some of the key parameters in the model fits are correlated (such as $\Delta a$ and $e_p$ in the $\sigma_{e_p}=0$ model) or have non-Gaussian posterior distributions, and so the median parameter values selected for these models may not truly be the ``best-fit" ones. 

One subtle detail in the model profiles that include a dispersion parameter is that the flux falls off more gently resulting in wider low-surface brightness features. This could at least partially explain the ``shoulders" seen in the data, and may also be the reason this model is statistically preferred over the other one despite the fact that it cannot reproduce the brightness asymmetry or width difference. These ``shoulders" appear slightly more distinct from the central peak of emission compared to the more smooth profile from the model, which may indicate that these are separate features. Edge sharpness can be used to constrain the properties of sculpting planets \cite[e.g.,][]{pearce:2024}.  The fact that the preferred model has smoother edges could also imply that other dynamical processes might be involved in creating the eccentricity in the Fomalhaut disk such as self-gravitation, particle collisions, and close-packing seen in planetary rings in the Solar System \citep{dermott:1980}. More detailed N-body simulations are needed to fully explore this.  The exterior ``shoulders" are also reminiscent of the halos detected in other debris disks such as HD 32297, HD 61005, and q$^1$ Eri \citep{MacGregor2018, Lovell2021}, and targets from the ARKS survey (in prep., private communication). The differences in the shape of the radial profiles at the two ansae, including the exterior shoulders, appear consistent with an N-body simulation of an eccentric planet sculpting an exterior disk presented in Appendix B1 of \citet{pearce:2024}, with a narrower apocenter and an exterior halo/shoulder, though the SE ansa profile of Fomalhaut appears more Gaussian than in that work. Higher resolution observations in the future may be able to resolve just how distinct these features are.

Flux from the intermediate belt in the JWST MIRI image and the extended exterior halo complicate the width measurement, and so we instead cite a lower limit by measuring the width between where the flux radially outwards falls to the mean flux level about 25 AU interior to the peak emission. In addition, the MIRI beam size is $\sim1\arcsec$ or $\sim7.7$~au, much larger than the ALMA beam. We place upper limits of 51 AU and 31 AU on the SE and NW ansae in the JWST image, or a width difference of about 20 AU between the ansae. However, if we instead compared where the flux level at the outer edge of the SE ansa is comparable to the flux beyond the dimmer NW ansa, the SE  width could be as much as 20 AU wider than estimated here. These estimates are better compared to the full width at 20$\%$ maximum measurements of 29 AU and 21 AU at the SE and NW ansae from the ALMA data (Section \ref{sec:imageanalysis}). 

Lastly, we note that relatively high inclination of the disk degrades our ability to accurately measure the deprojected disk width away from the major axis (and is compounded by the increase in noise and artifacts in these same regions as discussed in Section \ref{sec:imaging}).

\subsection{Flux Constraints on the Intermediate Belt}
\label{sec:ib}
We do not detect the Intermediate Belt (IB) revealed by JWST/MIRI \citep{Gaspar2023} in the ALMA mosaic at 1.3~mm.  In order to constrain the total flux for the IB at 1.3~mm, we estimate boundaries for the IB using parameters from \citet{Gaspar2023} and our forced and proper eccentricity model. \citet{Gaspar2023} note that the orbital boundaries of the IB are not well-defined due to the bright inner belt and the fainter NE ansa. They used ellipse fitting to estimate an inner boundary of $a=83$~AU, $e=0.31$, and an outer boundary of $a=104$~AU, $e=0.265$. They also found that the IB and inner belts had slightly varying inclinations and position angles, but for the sake of simplicity we ignore those differences here. Instead, we assume that the IB has a similar forced argument of pericenter as the outer belt, $\omega_f=45^{\circ}$, and trace out the orbits for the boundaries corresponding to all mean longitudes, $l = [0, 2\pi]$. These boundaries are plotted as white dashed lines in Figures \ref{fig:mosaic}c and d.

The approximate RMS noise and artifacts in the full mosaic image (Figure \ref{fig:mosaic}c) increase away from the pointing center at the disk ansae as discussed in Section \ref{sec:imaging}. To measure this effect, we create an annulus region in \texttt{CASA} defined by the above boundaries, and use the \texttt{imstat} task to compute the RMS, about 12 $\mu$Jy/bm. We therefore put a 3-$\sigma$ upper limit on the peak flux of the IB at 1.33~mm of 36~$\mu$Jy/bm. The area denoted by these approximate boundaries corresponds to $n_{\rm beams}=121$.  We then use the expression $F_{\mathrm{lim}}= (n_{\mathrm{beams}}\times \mathrm{rms})/\sqrt{n_{\mathrm{beams}}}$ to place a 3-$\sigma$ upper limit of 396~$\mu$Jy on the IB flux at 1.33~mm. Using these same boundaries, we compute 56~mJy of flux in the JWST 25.5~$\mu$m image, although we note that this estimate is likely contaminated with flux from the inner belt.

\citet{Sommer2025} suggest the IB could be explained by P-R drag along with unmodeled features, like  resonant trapping of small grains, without necessarily invoking a second, collisionally-active dust belt. Future SED modeling could use the upper limits these observations provide to clarify the nature of the IB.

\subsection{Great Dust Cloud Detection}
\label{sec:gdc}

\begin{figure*}
    \centering
    \includegraphics[width=\textwidth]{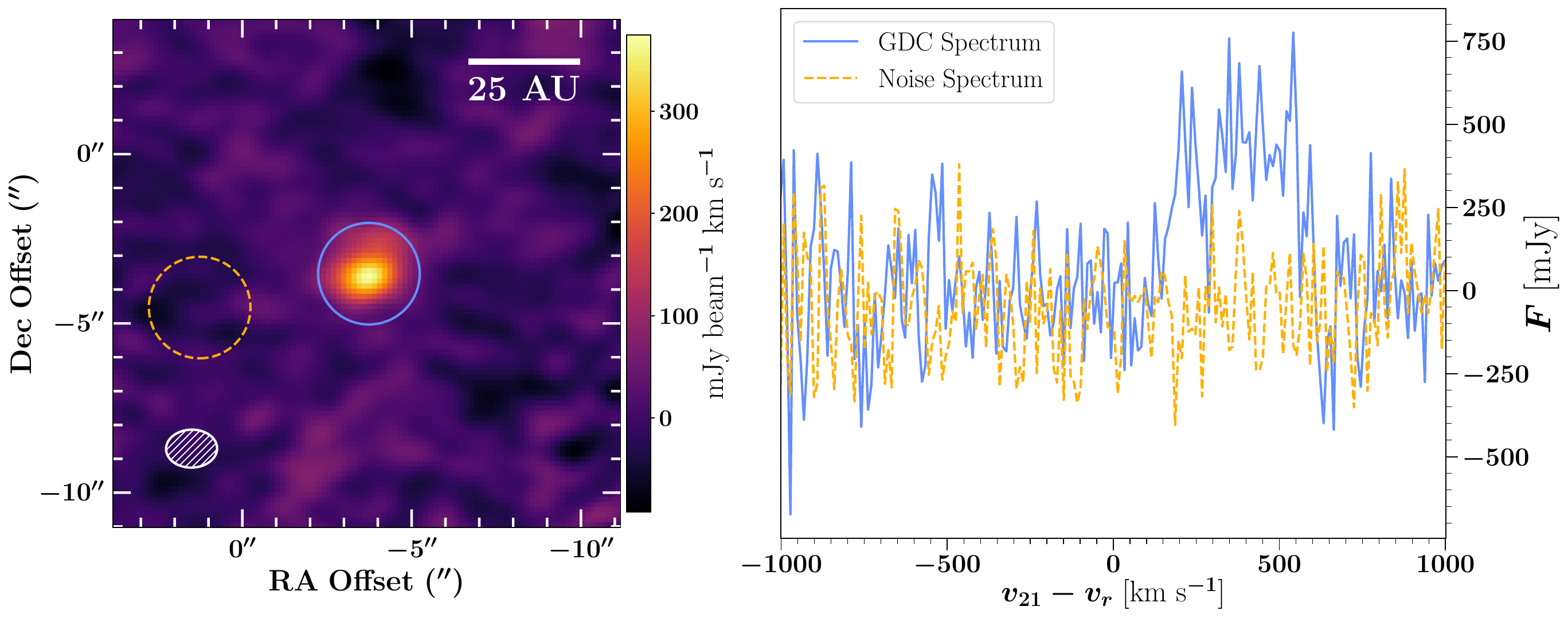}
    \caption{\textit{Left:} A Moment-0 image of the ``Great Dust Cloud" obtained by integrating between 230.1 and 230.4 GHz (173 to 563 km s$^{-1}$ channels relative to the expected CO line) of the Cycle 3 residual measurement set (i.e. the best-fit disk continuum model has been subtracted off). The cyan circle denotes the area of the region used to extract the spectrum of the source, and the orange-dashed circle is denotes a region of equivalent size with characteristic noise. \textit{Right:} The resulting extracted spectra of the background galaxy and noise region. The source emission is redshifted by $>$400 km s$^{-1}$ from the expected Doppler-shifted transition of $^{12}$CO $J = 2 - 1$  corresponding to Fomalhaut \citep[$v_r=6.5$ km s$^{-1}$,][]{Gontcharov2006}, suggesting the source is unaffiliated with the system. The width of the emission is $\approx$300 km s$^{-1}$, consistent with velocity dispersion from rotating disk galaxies.}
    \label{fig:gdc}
\end{figure*}

The spectral window corresponding to the C5 data centered on the $\nu_{21}=230.538$~GHz CO 2--1 line of the residual measurement set (i.e., the best-fit $\sigma_{e_p}$ continuum model was subtracted off) was imaged as a cube with \texttt{tclean}, and a Keplerian spatio-spectral mask was used in order to determine if CO emission from the exocometary debris could be detected as in \citet{Matra2017}. No emission was detected, but this result is consistent with the previous work given the higher resolution here.

However, upon reexamination of the disk model-subtracted C3 CO data, we discovered multiple consecutive channels of emission coincident with the position of the ``Great Dust Cloud" object discussed in \citet{Gaspar2023,Kennedy2023}. The emission occurs between 230.1 and 230.4 GHz, corresponding to a linewidth of $\approx$ 300~km~s$^{-1}$. Fomalhaut's systemic radial velocity is $v_r=6.5$~km~s$^{-1}$ \citep{Gontcharov2006} which would imply that the emitting gas is moving with velocities $>400$~km~s$^{-1}$ if they were from the expected 230~GHz CO 2--1 line. Given the dynamics of debris disks and the relative size, velocity, and location of this feature, it seems implausible that it is connected to a feature in the Fomalhaut debris disk and is much more likely to be yet-unidentified spectral line emission from a background galaxy.  In Figure \ref{fig:gdc}, we produce a Moment-0 image of the spectral line from a primary beam-corrected cube, and use two spatial filters to extract 1D source and noise spectra. No other sources appear in these channels above the noise level. The source was not detected in the C5 spectral window containing the $^{12}$CO (2-1) transition, likely because this location corresponds to below the 20$\%$ gain level of the primary beam and thus may be lost in the noise.

\citet{Ygouf2024} fit the available photometry of the galaxy using template spectra of galaxies at $z=$ 0.80, 0.21, and 0.56 and concluded that the GDC was an ultra-luminous IR galaxy. On the assumption that this line emanates from CO (plausibly the brightest line and so most likely to be detected), we consider one possibility that the spectral line we identified is the $^{12}$CO (3-2) transition at $\nu_{32}=345.796$ GHz. Given that the observed central frequency of the line is at $\nu_{\rm obs}\approx230.25$ GHz, we estimate a corresponding redshift of $z=0.502$. It is unclear how uncertain the redshift estimates from the fits presented in \citet{Ygouf2024} are, but this could support the $z=0.56$, NGC6240-analog they proposed.

\section{Conclusions}
\label{sec:conclusions}

We used long-baseline observations from ALMA Cycle 5 and short-baseline observations from Cycle 3 to produce a 0\farcs57 (4.4 AU) image of the Fomalhaut outer debris disk at a sensitivity of 7 $\mu$Jy/bm.

\begin{enumerate}

    \item We generated radial profiles of the new ALMA image and the 25.5 $\mu$m JWST/MIRI image from \citet{Gaspar2023} by regridding them to an $R-\theta$ grid. We then measured the FWHM of the radial profiles, and found that the SE side of the disk near pericenter is 4 AU wider than at the NW side near apocenter with ALMA, and about 20 AU wider with JWST. We also observed a $21\pm1\%$ brighter NW side compared to the SE side in the ALMA image from the peak brightness of the radial profiles.

    \item We performed MCMC fits of two, particle-based disk models with proper and forced component eccentricities to the ALMA visibilities. Our BIC analysis suggests that the model that includes a $\sigma_{e_p}$ parameter is preferred over the model without one, supporting the findings in \citet{Kennedy2020}.

    \item Neither model is able to replicate the 4 AU width difference, and instead have a 1 AU wider apocenter. Future modeling should test whether a Gaussian semi-major axis distribution for the particles as opposed to the uniform one employed here can alleviate this tension.

    \item Neither model could reproduce the $21\pm1\%$ brightness asymmetry near apocenter measured from the radial profile of the data, suggesting there is a yet missing component to the physics underlying the distribution of material in the disk. \citet{Lovell2025} present a model with an eccentricity gradient that can simultaneously account for the width and brightness difference.

    \item We do not detected any emission from the Intermediate Belt discovered in JWST/MIRI imaging, but are able to place a $3-\sigma$ upper limit of 396 $\mu$Jy for the total flux at 1.33 mm.

    \item We discovered a spectral line in the archival Cycle 3 data centered at $\nu_{\rm obs}\approx230.25$ GHz at the location of the ``Great Dust Cloud," redshifted from the expected CO (2-1) line by more than 400 km s$^{-1}$. This high velocity supports the conclusion that the object is a background galaxy. We suggest that the spectral line could be from CO (3-2) emission at a rest frequency of 345 GHz, implying a redshift of $z\approx0.502$.

\end{enumerate}

Fomalhaut's proximity allows observations to resolve its structure at higher resolution than other systems, which continues to make it an ideal target to explore the early evolution of planetary systems.  These new data reveal variation in the azimuthal structure of the outer disk that is not well-fit by current eccentric models.  Future observations could provide further insights into the disk's radial substructure on spatial scales where we could measure its eccentric morphology and test whether it is consistent with the architecture characteristic of sculpting by an internal planet.

\section*{Acknowledgments}
We acknowledge the contributions of Wayne Holland to the initial ALMA proposal that resulted in this paper.  Dr. Holland unfortunately passed away before this work reached its conclusion.

This paper makes use of the following ALMA data: ADS/JAO.ALMA\#2013.1.00486.S, \#2015.1.00966.S, and \#2017.1.01043.S. ALMA is a partnership of ESO (representing its member states), NSF (USA) and NINS (Japan), together with NRC (Canada), MOST and ASIAA (Taiwan), and KASI (Republic of Korea), in cooperation with the Republic of Chile. The Joint ALMA Observatory is operated by ESO, AUI/NRAO and NAOJ. The National Radio Astronomy Observatory is a facility of the National Science Foundation operated under cooperative agreement by Associated Universities, Inc.

This work utilized the Alpine high performance computing resource at the University of Colorado Boulder. Alpine is jointly funded by the University of Colorado Boulder, the University of Colorado Anschutz, Colorado State University, and the National Science Foundation (award 2201538).

This work utilized the Blanca condo computing resource at the University of Colorado Boulder. Blanca is jointly funded by computing users and the University of Colorado Boulder.

MAM acknowledges support for part of this research from the National Aeronautics and Space Administration (NASA) under award number 19-ICAR19\_2-0041.

JBL acknowledges the Smithsonian Institute for funding via a Submillimeter Array (\textit{SMA}) Fellowship.

JSC gratefully acknowledges Kirk Long for optimizing the particle-based models presented in this work.

LM acknowledges funding by the European Union through the E-BEANS ERC project (grant number 100117693). Views and opinions expressed are however those of the author(s) only and do not necessarily reflect those of the European Union or the European Research Council Executive Agency. Neither the European Union nor the granting authority can be held responsible for them.

OP acknowledges support from the Science
and Technology Facilities Council of the United Kingdom, grant number ST/T000287/1.

AMH gratefully acknowledges support from the National Science Foundation under Grant No. ASTR-2307920.

\software{\texttt{numpy} \citep{numpy}, \texttt{matplotlib} \citep{matplotlib}, \texttt{CASA} \citep{McMullin2007}, \texttt{uvmodelfit} \citep{Marti-Vidal2014}, \texttt{astropy}, \texttt{regions}, \texttt{spectral-cube}, \citep{astropy5}, \texttt{emcee} \citep{ForemanMackey2013}, \texttt{galario} \citep{Tazzari2018}, \texttt{kepler} \citep{ForemanMackey2021}, \texttt{scipy} \citep{scipy}, \texttt{numba} \citep{numba}}

\bibliography{bib.bib}

\begin{thebibliography}{}
\expandafter\ifx\csname natexlab\endcsname\relax\def\natexlab#1{#1}\fi
\providecommand{\url}[1]{\href{#1}{#1}}
\providecommand{\dodoi}[1]{doi:~\href{http://doi.org/#1}{\nolinkurl{#1}}}
\providecommand{\doeprint}[1]{\href{http://ascl.net/#1}{\nolinkurl{http://ascl.net/#1}}}
\providecommand{\doarXiv}[1]{\href{https://arxiv.org/abs/#1}{\nolinkurl{https://arxiv.org/abs/#1}}}

\bibitem[{{Acke} {et~al.}(2012){Acke}, {Min}, {Dominik}, {Vandenbussche},
  {Sibthorpe}, {Waelkens}, {Olofsson}, {Degroote}, {Smolders}, {Pantin},
  {Barlow}, {Blommaert}, {Brandeker}, {De Meester}, {Dent}, {Exter}, {Di
  Francesco}, {Fridlund}, {Gear}, {Glauser}, {Greaves}, {Harvey}, {Henning},
  {Hogerheijde}, {Holland}, {Huygen}, {Ivison}, {Jean}, {Liseau}, {Naylor},
  {Pilbratt}, {Polehampton}, {Regibo}, {Royer}, {Sicilia-Aguilar}, \&
  {Swinyard}}]{Acke2012}
{Acke}, B., {Min}, M., {Dominik}, C., {et~al.} 2012, \aap, 540, A125,
  \dodoi{10.1051/0004-6361/201118581}

\bibitem[{{Astropy Collaboration} {et~al.}(2022){Astropy Collaboration},
  {Price-Whelan}, {Lim}, {Earl}, {Starkman}, {Bradley}, {Shupe}, {Patil},
  {Corrales}, {Brasseur}, {N{\"o}the}, {Donath}, {Tollerud}, {Morris},
  {Ginsburg}, {Vaher}, {Weaver}, {Tocknell}, {Jamieson}, {van Kerkwijk},
  {Robitaille}, {Merry}, {Bachetti}, {G{\"u}nther}, {Aldcroft},
  {Alvarado-Montes}, {Archibald}, {B{\'o}di}, {Bapat}, {Barentsen},
  {Baz{\'a}n}, {Biswas}, {Boquien}, {Burke}, {Cara}, {Cara}, {Conroy},
  {Conseil}, {Craig}, {Cross}, {Cruz}, {D'Eugenio}, {Dencheva}, {Devillepoix},
  {Dietrich}, {Eigenbrot}, {Erben}, {Ferreira}, {Foreman-Mackey}, {Fox},
  {Freij}, {Garg}, {Geda}, {Glattly}, {Gondhalekar}, {Gordon}, {Grant},
  {Greenfield}, {Groener}, {Guest}, {Gurovich}, {Handberg}, {Hart},
  {Hatfield-Dodds}, {Homeier}, {Hosseinzadeh}, {Jenness}, {Jones}, {Joseph},
  {Kalmbach}, {Karamehmetoglu}, {Ka{\l}uszy{\'n}ski}, {Kelley}, {Kern},
  {Kerzendorf}, {Koch}, {Kulumani}, {Lee}, {Ly}, {Ma}, {MacBride}, {Maljaars},
  {Muna}, {Murphy}, {Norman}, {O'Steen}, {Oman}, {Pacifici}, {Pascual},
  {Pascual-Granado}, {Patil}, {Perren}, {Pickering}, {Rastogi}, {Roulston},
  {Ryan}, {Rykoff}, {Sabater}, {Sakurikar}, {Salgado}, {Sanghi}, {Saunders},
  {Savchenko}, {Schwardt}, {Seifert-Eckert}, {Shih}, {Jain}, {Shukla}, {Sick},
  {Simpson}, {Singanamalla}, {Singer}, {Singhal}, {Sinha}, {Sip{\H{o}}cz},
  {Spitler}, {Stansby}, {Streicher}, {{\v{S}}umak}, {Swinbank}, {Taranu},
  {Tewary}, {Tremblay}, {de Val-Borro}, {Van Kooten}, {Vasovi{\'c}}, {Verma},
  {de Miranda Cardoso}, {Williams}, {Wilson}, {Winkel}, {Wood-Vasey}, {Xue},
  {Yoachim}, {Zhang}, {Zonca}, \& {Astropy Project Contributors}}]{astropy5}
{Astropy Collaboration}, {Price-Whelan}, A.~M., {Lim}, P.~L., {et~al.} 2022,
  \apj, 935, 167, \dodoi{10.3847/1538-4357/ac7c74}

\bibitem[{{Backman} \& {Paresce}(1993)}]{Backman1993}
{Backman}, D.~E., \& {Paresce}, F. 1993, in Protostars and Planets III, ed.
  E.~H. {Levy} \& J.~I. {Lunine}, 1253

\bibitem[{{Boley} {et~al.}(2012){Boley}, {Payne}, {Corder}, {Dent}, {Ford}, \&
  {Shabram}}]{Boley2012}
{Boley}, A.~C., {Payne}, M.~J., {Corder}, S., {et~al.} 2012, \apjl, 750, L21,
  \dodoi{10.1088/2041-8205/750/1/L21}

\bibitem[{{Dermott} \& {Murray}(1980)}]{dermott:1980}
{Dermott}, S.~F., \& {Murray}, C.~D. 1980, \icarus, 43, 338,
  \dodoi{10.1016/0019-1035(80)90179-7}

\bibitem[{{Foreman-Mackey} {et~al.}(2013){Foreman-Mackey}, {Hogg}, {Lang}, \&
  {Goodman}}]{ForemanMackey2013}
{Foreman-Mackey}, D., {Hogg}, D.~W., {Lang}, D., \& {Goodman}, J. 2013, \pasp,
  125, 306, \dodoi{10.1086/670067}

\bibitem[{{Foreman-Mackey} {et~al.}(2021){Foreman-Mackey}, {Luger}, {Agol},
  {Barclay}, {Bouma}, {Brandt}, {Czekala}, {David}, {Dong}, {Gilbert},
  {Gordon}, {Hedges}, {Hey}, {Morris}, {Price-Whelan}, \&
  {Savel}}]{ForemanMackey2021}
{Foreman-Mackey}, D., {Luger}, R., {Agol}, E., {et~al.} 2021, arXiv e-prints,
  arXiv:2105.01994.
\newblock \doarXiv{2105.01994}

\bibitem[{{G{\'a}sp{\'a}r} \& {Rieke}(2020)}]{Gaspar2020}
{G{\'a}sp{\'a}r}, A., \& {Rieke}, G. 2020, Proceedings of the National Academy
  of Science, 117, 9712, \dodoi{10.1073/pnas.1912506117}

\bibitem[{{G{\'a}sp{\'a}r} {et~al.}(2023){G{\'a}sp{\'a}r}, {Wolff}, {Rieke},
  {Leisenring}, {Morrison}, {Su}, {Ward-Duong}, {Aguilar}, {Ygouf}, {Beichman},
  {Llop-Sayson}, \& {Bryden}}]{Gaspar2023}
{G{\'a}sp{\'a}r}, A., {Wolff}, S.~G., {Rieke}, G.~H., {et~al.} 2023, Nature
  Astronomy, 7, 790, \dodoi{10.1038/s41550-023-01962-6}

\bibitem[{{Gontcharov}(2006)}]{Gontcharov2006}
{Gontcharov}, G.~A. 2006, Astronomy Letters, 32, 759,
  \dodoi{10.1134/S1063773706110065}

\bibitem[{Harris {et~al.}(2020)Harris, Millman, van~der Walt, Gommers,
  Virtanen, Cournapeau, Wieser, Taylor, Berg, Smith, Kern, Picus, Hoyer, van
  Kerkwijk, Brett, Haldane, Fernández~del Río, Wiebe, Peterson,
  Gérard-Marchant, Sheppard, Reddy, Weckesser, Abbasi, Gohlke, \&
  Oliphant}]{numpy}
Harris, C.~R., Millman, K.~J., van~der Walt, S.~J., {et~al.} 2020, Nature, 585,
  357–362, \dodoi{10.1038/s41586-020-2649-2}

\bibitem[{{Holland} {et~al.}(1998){Holland}, {Greaves}, {Zuckerman}, {Webb},
  {McCarthy}, {Coulson}, {Walther}, {Dent}, {Gear}, \& {Robson}}]{Holland1998}
{Holland}, W.~S., {Greaves}, J.~S., {Zuckerman}, B., {et~al.} 1998, \nat, 392,
  788, \dodoi{10.1038/33874}

\bibitem[{Hunter(2007)}]{matplotlib}
Hunter, J.~D. 2007, Computing in Science \& Engineering, 9, 90,
  \dodoi{10.1109/MCSE.2007.55}

\bibitem[{{Kalas} {et~al.}(2005){Kalas}, {Graham}, \& {Clampin}}]{Kalas2005}
{Kalas}, P., {Graham}, J.~R., \& {Clampin}, M. 2005, \nat, 435, 1067,
  \dodoi{10.1038/nature03601}

\bibitem[{{Kalas} {et~al.}(2013){Kalas}, {Graham}, {Fitzgerald}, \&
  {Clampin}}]{Kalas2013}
{Kalas}, P., {Graham}, J.~R., {Fitzgerald}, M.~P., \& {Clampin}, M. 2013, \apj,
  775, 56, \dodoi{10.1088/0004-637X/775/1/56}

\bibitem[{{Kalas} {et~al.}(2008){Kalas}, {Graham}, {Chiang}, {Fitzgerald},
  {Clampin}, {Kite}, {Stapelfeldt}, {Marois}, \& {Krist}}]{Kalas2008}
{Kalas}, P., {Graham}, J.~R., {Chiang}, E., {et~al.} 2008, Science, 322, 1345,
  \dodoi{10.1126/science.1166609}

\bibitem[{Kass \& Raftery(1995)}]{Kass1995}
Kass, R.~E., \& Raftery, A.~E. 1995, Journal of the American Statistical
  Association, 90, 773, \dodoi{10.1080/01621459.1995.10476572}

\bibitem[{{Kennedy}(2020)}]{Kennedy2020}
{Kennedy}, G.~M. 2020, Royal Society Open Science, 7, 200063,
  \dodoi{10.1098/rsos.200063}

\bibitem[{{Kennedy} {et~al.}(2023){Kennedy}, {Lovell}, {Kalas}, \&
  {Fitzgerald}}]{Kennedy2023}
{Kennedy}, G.~M., {Lovell}, J.~B., {Kalas}, P., \& {Fitzgerald}, M.~P. 2023,
  \mnras, 524, 2698, \dodoi{10.1093/mnras/stad2058}

\bibitem[{{Kenyon} {et~al.}(2014){Kenyon}, {Currie}, \& {Bromley}}]{Kenyon2014}
{Kenyon}, S.~J., {Currie}, T., \& {Bromley}, B.~C. 2014, \apj, 786, 70,
  \dodoi{10.1088/0004-637X/786/1/70}

\bibitem[{Lam {et~al.}(2015)Lam, Pitrou, \& Seibert}]{numba}
Lam, S.~K., Pitrou, A., \& Seibert, S. 2015, in Proceedings of the Second
  Workshop on the LLVM Compiler Infrastructure in HPC, LLVM '15 (New York, NY,
  USA: Association for Computing Machinery), \dodoi{10.1145/2833157.2833162}

\bibitem[{{Lawler} {et~al.}(2015){Lawler}, {Greenstreet}, \&
  {Gladman}}]{Lawler2015}
{Lawler}, S.~M., {Greenstreet}, S., \& {Gladman}, B. 2015, \apjl, 802, L20,
  \dodoi{10.1088/2041-8205/802/2/L20}

\bibitem[{{Lovell} \& {Lynch}(2023)}]{Lovell2023}
{Lovell}, J.~B., \& {Lynch}, E.~M. 2023, \mnras, 525, L36,
  \dodoi{10.1093/mnrasl/slad083}

\bibitem[{{Lovell} {et~al.}(2021){Lovell}, {Marino}, {Wyatt}, {Kennedy},
  {MacGregor}, {Stapelfeldt}, {Dent}, {Krist}, {Matr{\`a}}, {Kral},
  {Pani{\'c}}, {Pearce}, \& {Wilner}}]{Lovell2021}
{Lovell}, J.~B., {Marino}, S., {Wyatt}, M.~C., {et~al.} 2021, \mnras, 506,
  1978, \dodoi{10.1093/mnras/stab1678}

\bibitem[{{Lovell} {et~al.}(2025){Lovell}, {Lynch}, Chittidi, Sefilian,
  Andrews, Kennedy, MacGregor, Wilner, \& Wyatt}]{Lovell2025}
{Lovell}, J.~B., {Lynch}, E.~M., Chittidi, J.~S., {et~al.} 2025, \apj,
  \dodoi{10.3847/1538-4357/adfadc}

\bibitem[{{Lynch} \& {Lovell}(2022)}]{Lynch2022}
{Lynch}, E.~M., \& {Lovell}, J.~B. 2022, \mnras, 510, 2538,
  \dodoi{10.1093/mnras/stab3566}

\bibitem[{{MacGregor} {et~al.}(2017){MacGregor}, {Matr{\`a}}, {Kalas},
  {Wilner}, {Pan}, {Kennedy}, {Wyatt}, {Duchene}, {Hughes}, {Rieke}, {Clampin},
  {Fitzgerald}, {Graham}, {Holland}, {Pani{\'c}}, {Shannon}, \&
  {Su}}]{MacGregor2017}
{MacGregor}, M.~A., {Matr{\`a}}, L., {Kalas}, P., {et~al.} 2017, \apj, 842, 8,
  \dodoi{10.3847/1538-4357/aa71ae}

\bibitem[{{MacGregor} {et~al.}(2018){MacGregor}, {Weinberger}, {Hughes},
  {Wilner}, {Currie}, {Debes}, {Donaldson}, {Redfield}, {Roberge}, \&
  {Schneider}}]{MacGregor2018}
{MacGregor}, M.~A., {Weinberger}, A.~J., {Hughes}, A.~M., {et~al.} 2018, \apj,
  869, 75, \dodoi{10.3847/1538-4357/aaec71}

\bibitem[{{MacGregor} {et~al.}(2022){MacGregor}, {Hurt}, {Stark}, {Howard},
  {Weinberger}, {Ren}, {Schneider}, {Choquet}, \& {Mawet}}]{MacGregor2022}
{MacGregor}, M.~A., {Hurt}, S.~A., {Stark}, C.~C., {et~al.} 2022, \apjl, 933,
  L1, \dodoi{10.3847/2041-8213/ac7729}

\bibitem[{{Mamajek}(2012)}]{Mamajek2012}
{Mamajek}, E.~E. 2012, \apjl, 754, L20, \dodoi{10.1088/2041-8205/754/2/L20}

\bibitem[{{Marino}(2019)}]{Marino2019}
{Marino}, S. 2019, \apj, 881, 84, \dodoi{10.3847/1538-4357/ab2b98}

\bibitem[{{Marino}(2022)}]{Marino2022}
---. 2022, arXiv e-prints, arXiv:2202.03053, \dodoi{10.48550/arXiv.2202.03053}

\bibitem[{{Mart{\'\i}-Vidal} {et~al.}(2014){Mart{\'\i}-Vidal}, {Vlemmings},
  {Muller}, \& {Casey}}]{Marti-Vidal2014}
{Mart{\'\i}-Vidal}, I., {Vlemmings}, W.~H.~T., {Muller}, S., \& {Casey}, S.
  2014, \aap, 563, A136, \dodoi{10.1051/0004-6361/201322633}

\bibitem[{{Matr{\`a}} {et~al.}(2017){Matr{\`a}}, {MacGregor}, {Kalas}, {Wyatt},
  {Kennedy}, {Wilner}, {Duchene}, {Hughes}, {Pan}, {Shannon}, {Clampin},
  {Fitzgerald}, {Graham}, {Holland}, {Pani{\'c}}, \& {Su}}]{Matra2017}
{Matr{\`a}}, L., {MacGregor}, M.~A., {Kalas}, P., {et~al.} 2017, \apj, 842, 9,
  \dodoi{10.3847/1538-4357/aa71b4}

\bibitem[{{McMullin} {et~al.}(2007){McMullin}, {Waters}, {Schiebel}, {Young},
  \& {Golap}}]{McMullin2007}
{McMullin}, J.~P., {Waters}, B., {Schiebel}, D., {Young}, W., \& {Golap}, K.
  2007, in Astronomical Society of the Pacific Conference Series, Vol. 376,
  Astronomical Data Analysis Software and Systems XVI, ed. R.~A. {Shaw},
  F.~{Hill}, \& D.~J. {Bell}, 127

\bibitem[{{Pan} {et~al.}(2016){Pan}, {Nesvold}, \& {Kuchner}}]{Pan2016}
{Pan}, M., {Nesvold}, E.~R., \& {Kuchner}, M.~J. 2016, \apj, 832, 81,
  \dodoi{10.3847/0004-637X/832/1/81}

\bibitem[{{Pearce} {et~al.}(2024){Pearce}, {Krivov}, {Sefilian}, {Jankovic},
  {L{\"o}hne}, {Morgner}, {Wyatt}, {Booth}, \& {Marino}}]{pearce:2024}
{Pearce}, T.~D., {Krivov}, A.~V., {Sefilian}, A.~A., {et~al.} 2024, \mnras,
  527, 3876, \dodoi{10.1093/mnras/stad3462}

\bibitem[{{Sepulveda} {et~al.}(2019){Sepulveda}, {Matr{\`a}}, {Kennedy}, {del
  Burgo}, {{\"O}berg}, {Wilner}, {Marino}, {Booth}, {Carpenter}, {Davies},
  {Dent}, {Ertel}, {Lestrade}, {Marshall}, {Milli}, {Wyatt}, {MacGregor}, \&
  {Matthews}}]{Sepulveda2019}
{Sepulveda}, A.~G., {Matr{\`a}}, L., {Kennedy}, G.~M., {et~al.} 2019, \apj,
  881, 84, \dodoi{10.3847/1538-4357/ab2b98}

\bibitem[{{Sommer} {et~al.}(2025){Sommer}, {Wyatt}, \& {Han}}]{Sommer2025}
{Sommer}, M., {Wyatt}, M., \& {Han}, Y. 2025, \mnras,
  \dodoi{10.1093/mnras/staf494}

\bibitem[{{Stapelfeldt} {et~al.}(2004){Stapelfeldt}, {Holmes}, {Chen}, {Rieke},
  {Su}, {Hines}, {Werner}, {Beichman}, {Jura}, {Padgett}, {Stansberry},
  {Bendo}, {Cadien}, {Marengo}, {Thompson}, {Velusamy}, {Backus}, {Blaylock},
  {Egami}, {Engelbracht}, {Frayer}, {Gordon}, {Keene}, {Latter}, {Megeath},
  {Misselt}, {Morrison}, {Muzerolle}, {Noriega-Crespo}, {Van Cleve}, \&
  {Young}}]{Stapelfeldt2004}
{Stapelfeldt}, K.~R., {Holmes}, E.~K., {Chen}, C., {et~al.} 2004, \apjs, 154,
  458, \dodoi{10.1086/423135}

\bibitem[{{Tazzari} {et~al.}(2018){Tazzari}, {Beaujean}, \&
  {Testi}}]{Tazzari2018}
{Tazzari}, M., {Beaujean}, F., \& {Testi}, L. 2018, \mnras, 476, 4527,
  \dodoi{10.1093/mnras/sty409}

\bibitem[{{van Leeuwen}(2007)}]{vanLeeuwen2007}
{van Leeuwen}, F. 2007, \aap, 474, 653, \dodoi{10.1051/0004-6361:20078357}

\bibitem[{Virtanen {et~al.}(2020)Virtanen, Gommers, Oliphant, Haberland, Reddy,
  Cournapeau, Burovski, Peterson, Weckesser, Bright, {van der Walt}, Brett,
  Wilson, Millman, Mayorov, Nelson, Jones, Kern, Larson, Carey, Polat, Feng,
  Moore, {VanderPlas}, Laxalde, Perktold, Cimrman, Henriksen, Quintero, Harris,
  Archibald, Ribeiro, Pedregosa, {van Mulbregt}, \& {SciPy 1.0
  Contributors}}]{scipy}
Virtanen, P., Gommers, R., Oliphant, T.~E., {et~al.} 2020, Nature Methods, 17,
  261, \dodoi{10.1038/s41592-019-0686-2}

\bibitem[{{White} {et~al.}(2017){White}, {Boley}, {Dent}, {Ford}, \&
  {Corder}}]{White2017}
{White}, J.~A., {Boley}, A.~C., {Dent}, W.~R.~F., {Ford}, E.~B., \& {Corder},
  S. 2017, \mnras, 466, 4201, \dodoi{10.1093/mnras/stw3303}

\bibitem[{{Wyatt} {et~al.}(1999){Wyatt}, {Dermott}, {Telesco}, {Fisher},
  {Grogan}, {Holmes}, \& {Pi{\~n}a}}]{Wyatt1999}
{Wyatt}, M.~C., {Dermott}, S.~F., {Telesco}, C.~M., {et~al.} 1999, \apj, 527,
  918, \dodoi{10.1086/308093}

\bibitem[{{Ygouf} {et~al.}(2024){Ygouf}, {Beichman}, {Llop-Sayson}, {Bryden},
  {Leisenring}, {G{\'a}sp{\'a}r}, {Krist}, {Rieke}, {Rieke}, {Wolff},
  {Roellig}, {Su}, {Hainline}, {Hodapp}, {Greene}, {Meyer}, {Kelly}, {Misselt},
  {Stansberry}, {Boyer}, {Johnstone}, {Horner}, \& {Greenbaum}}]{Ygouf2024}
{Ygouf}, M., {Beichman}, C.~A., {Llop-Sayson}, J., {et~al.} 2024, \aj, 167, 26,
  \dodoi{10.3847/1538-3881/ad08c8}

\end{thebibliography}

\end{document}